\documentclass[aps,twocolumn,floatfix,superscriptaddress,prl]{revtex4-1}
\usepackage{amssymb}
\usepackage{graphicx}
\usepackage{dcolumn}
\usepackage{amsmath}
\usepackage{bm}
\usepackage{yhmath}
\usepackage{textcomp}
\usepackage{epstopdf}
\usepackage{color}
\usepackage{ulem}
\usepackage[toc,page,title,titletoc,header]{appendix}
\usepackage[colorlinks,
            linkcolor=blue,
            anchorcolor=blue,
            citecolor=blue
            ]{hyperref}
\usepackage{natbib}

\usepackage{tikz}
\newcommand*\emptycirc[1][0.4ex]{\tikz\draw (0,0) circle (#1);} 
\newcommand*\fullcirc[1][0.4ex]{\tikz\fill (0,0) circle (#1);}

\newcommand{\be}{\begin{eqnarray}}
\newcommand{\ee}{\end{eqnarray}}

\begin{document}

\title{Hilbert space shattering and disorder-free localization in polar lattice gases}

\author{Wei-Han Li}
\affiliation{Institute of Theoretical Physics, Leibniz University Hanover, Germany}
\author{Xiaolong Deng}
\affiliation{Institute of Theoretical Physics, Leibniz University Hanover, Germany}
\author{Luis Santos}
\affiliation{Institute of Theoretical Physics, Leibniz University Hanover, Germany}


\begin{abstract}
Emerging dynamical constraints resulting from inter-site interactions severely limit particle mobility in polar lattice gases. 
Whereas in absence of disorder hard-core Hubbard models with only strong nearest-neighbor interactions present Hilbert space fragmentation 
but no many-body localization for typical states, the $1/r^3$ tail of the dipolar interaction results in Hilbert space shattering, as well as in a dramatically slowed down dynamics and eventual 
disorder-free localization. Our results show that the study of the intriguing interplay between 
disorder- and interaction-induced many-body localization is within reach of future experiments with magnetic atoms and polar molecules.
\end{abstract}

\maketitle




Recent years have witnessed a considerable attention on the dynamics of many-body quantum systems, a 
rich topic both fundamentally and practically relevant~\cite{Eisert2014, DAlessio2016}.  Most quantum many-body systems are 
believed to thermalize as a consequence of the eigenstate thermalization hypothesis~\cite{Deutsch1991,Srednicki1994, Tasaki1998, Rigol2008}. 
Prominent exceptions to this paradigm include integrable systems~\cite{Rigol2007} 
 and many-body localization~(MBL) in disordered systems~\cite{Nandkishore2015, Altman2015, Abanin2019}. Progress on 
MBL has been recently followed by interest on MBL-like phenomenology in absence of disorder~\cite{Carleo2012,Grover2014,Schiulaz2015,Horssen2015, Barbiero2015,Papic2015,Hickey2016,Smith2017,Mondaini2017,Schulz2019, Nieuwenburg2019, Taylor2020, Chanda2020, Yao2020, Scherg2020, Guo2020, Morong2021, Yao2021}. 
Disorder-free localization occurs naturally due to dynamical constraints~\cite{Lan2018, Feldmeier2019, Nandkishore2019}.
These constraints, which result in a finite number of conservation laws, induce Hilbert space fragmentation into disconnected 
subspaces that severely limits the dynamics~\cite{DeTomasi2019,Pietracarpina2019,Moudgalya2019, Sala2020,Khemani2020, Herviou2020, Yang2020}. 
Hilbert space fragmentation is also closely connected to quantum scars~\cite{Turner2018}.


Ultra-cold gases in optical lattices or reconfigurable arrays provide a well-controlled scenario for the study of many-body dynamics,  
including MBL~\cite{Schreiber2015, Choi2016}, and quantum scars~\cite{Bernien2017}. 
Recent experiments on tilted Fermi-Hubbard chains~\cite{Scherg2020} and  in a trapped-ion quantum simulator~\cite{Morong2021} 
have provided evidence of non-ergodic behavior in absence of disorder, unveiling 
the potential of ultra-cold gases for the study of disorder-free MBL and Hilbert space fragmentation. It is hence particularly relevant to find other promising ultra-cold scenarios for the study of fragmentation due to interaction-induced constraints. As shown below, polar lattice gases are a natural candidate.


Power-law interacting systems have been the focus of recent breakthrough experiments, including trapped ions~\cite{Richerme2014, Jurcevic2014}, 
Rydberg gases~\cite{Bernien2017,Browaeys2016}, and lattice gases of magnetic atoms or polar molecules, with strong magnetic or 
electric dipole-dipole interactions. Experiments on polar lattice gases have already revealed 
inter-site spin-exchange in both atoms~\cite{DePaz2013} and molecules~\cite{Yan2013}, and realized an extended Hubbard model
with nearest-neighbor~(NN) interactions~\cite{Baier2016}. These experiments have started to unveil 
the fascinating possibilities that inter-site dipolar interactions offer 
for the quantum simulation of a large variety of models~\cite{Baranov2012}. 
MBL in disordered spin models with power-law Ising and exchange interactions has attracted a growing attention 
~\cite{Burin2006,Yao2014,Burin2015a,Burin2015b,Safavi2019,Roy2019,Schiffer2019,Deng2020}. Very recently,  
disorder-free Stark-MBL has been revealed in trapped ions with long-range spin-exchange~\cite{Morong2021}.
Inter-site interactions result as well in an intriguing dynamics in extended Hubbard models~\cite{Valiente2009,Nguenang2009,Petrosyan2007,Li2020,Morera2021, Fukuhara2013, Salerno2020, Li2020b}. 
In particular, NN dimers significantly slow-down the dynamics in one-dimensional~(1D) 
polar lattice gases, and may induce quasi-localization~\cite{Barbiero2015,Li2020}.



\begin{figure}[t]
\includegraphics[width=\columnwidth]{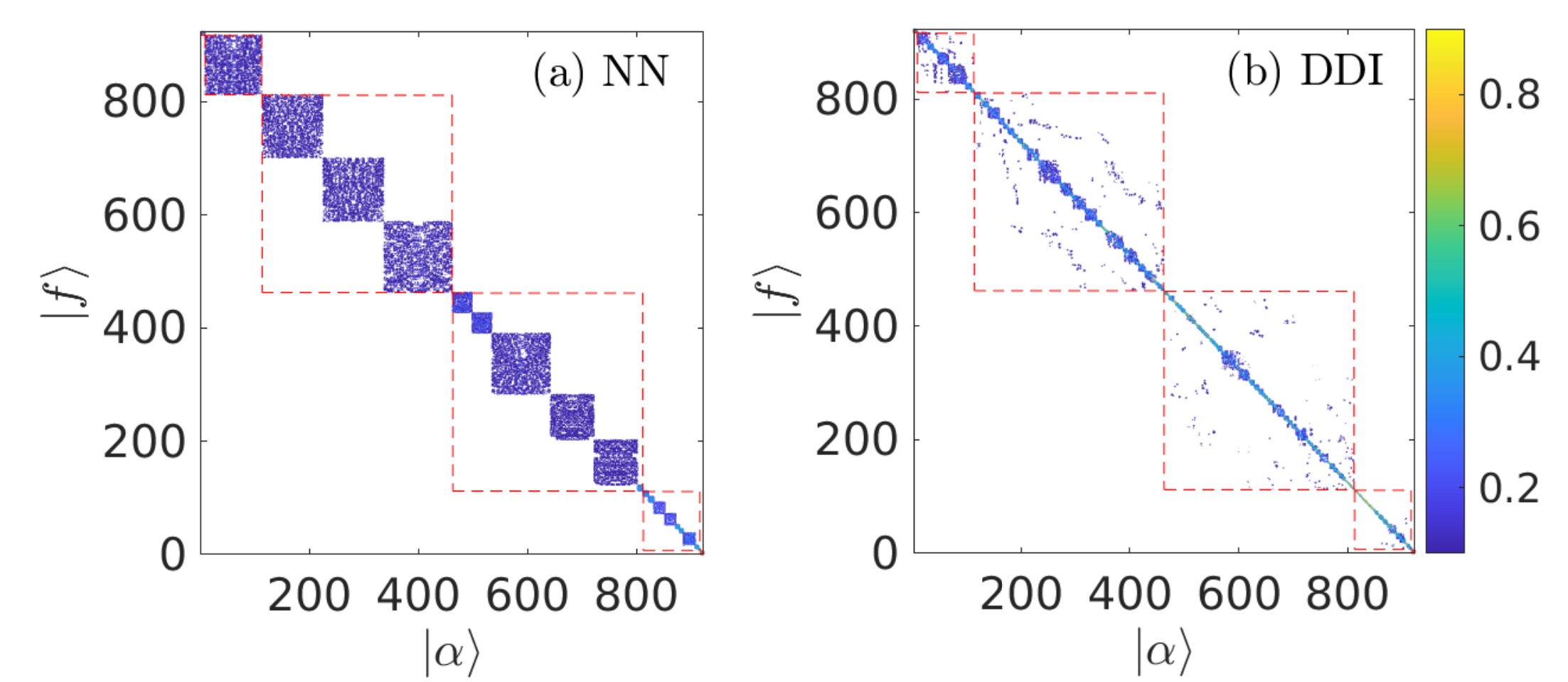} \centering
\caption{(Color online) Amplitude $|\psi_\alpha(f)|$ of the eigenstates $|\alpha\rangle$ in the Fock basis $\{|f\rangle\}$
for the NN model~(a) and the polar gas~(b) for $N=6$, $L=12$, $W=0$ and $V/t=50$. 
}
\label{fig:1}
\end{figure}



In the absence of any tilting or overall potential, a disorder-free system with only NN interactions (NN model) presents Hilbert space fragmentation due to the conservation of the number of NN bonds, 
but resonant motion within a fragment remains in general possible, precluding disorder-free localization~\cite{DeTomasi2019}. 
In this Letter we show that the $1/r^3$ tail of strong-enough dipolar interactions induces additional constraints that 
lead to the shattering~\cite{Khemani2020} of the Hilbert-space fragments of the NN model, and disrupt resonant motion within a Hilbert space fragment. 
The latter results in a very strong slow-down of the dynamics compared to the NN model, and eventually to disorder-free localization. 
Our results show that the study of disorder-free localization is within reach of future experiments on polar lattice gases. 


\paragraph{Model.--} We consider a 1D polar lattice gas of hard-core bosons~\cite{footnote-hardcore}
well described by the extended Hubbard model:
\begin{equation}
H=-t\sum_{j} \left( b^{\dagger}_{j} b_{j+1} + \mathrm{H.c.} \right) + \sum_{j} \epsilon_j n_j + \sum_{i < j} V_{ij} n_{i}n_{j},
\label{eq:H}
\end{equation}
where $V_{ij}=\frac{V}{|i-j|^{3}}$, $b_{j}(b^{\dagger}_{j})$ is the annihilation~(creation) operator at site $j$, $(b_j^\dag)^2=0$, 
$n_{j}=b^{\dagger}_{j}b_{j}$ the number operator, and $t$ the hopping amplitude. 
The random on-site energy $\epsilon_j$ is uniformly distributed in the interval $[-W,W]$.  
Our results are based on exact diagonalization of Eq.~\eqref{eq:H}.



\paragraph{Nearest-neighbor model.--} 
For the NN model, with 
$V_{ij}=V\delta_{j,i+1}$~\cite{BarLev2015, DeTomasi2019}, a dynamical constraint emerges for growing $V/t$, becoming 
exact for $V=\infty$, given by the conservation of NN bonds, $N_{\mathrm{NN}}=\sum_j \langle n_j n_{j+1}\rangle$. 
As a result, the Hilbert-space fragments into disconnected blocks of Fock states~\cite{DeTomasi2019}. However, crucially, the dynamics within each one of the blocks 
remains in general resonant, even for $V=\infty$.  For blocks with a finite density of singlons (i.e. particles without nearest neighbors), 
clusters of consecutively occupied sites of any length delocalize by swapping their positions with incoming singlons, through a series of resonant moves: 
$|\dots\emptycirc\,\emptycirc\,\fullcirc\,\fullcirc\dots \fullcirc\,\fullcirc\,\emptycirc\,\fullcirc\,\emptycirc\dots\rangle \to \dots \to |\dots\emptycirc\,\fullcirc\,\emptycirc\,\fullcirc\,\fullcirc\dots \fullcirc\,\fullcirc\,\emptycirc\dots\rangle$. 
As a result disorder-free localization is generally precluded~\cite{DeTomasi2019}, and delocalization occurs in a time $\sim 1/t$.



\paragraph{Hilbert space fragmentation.--} In order to study Hilbert-space fragmentation~\cite{Deng2021}, we 
obtain for $W=0$ the eigenstates $|\alpha\rangle = \sum_f \psi_\alpha(f) |f\rangle$, where $|f\rangle = \prod_{l=1}^L  |n_l (f)\rangle$ 
are the Fock states with population $n_l(f)=0,1$ in site $l$, and $N=\sum_{l=1}^L n_l$. 
Given an eigenstate $|\alpha\rangle$, we find the Fock states contributing to it~(up to a threshold $|\psi_\alpha(f)| > t^2/V$). 
We then determine the eigenstates with significant support on those Fock states, and iterate by proceeding similarly
with each of those eigenstates. Convergence is achieved after few iterations. For large-enough $V/t$, this procedure provides at $W=0$ the block of Fock states 
that connect with an initial $|f\rangle$ if allowed an infinitely long time.
Hilbert-space fragmentation is evident in Fig.~\ref{fig:1}(a)~(connected Fock states and eigenstates are bunched in consecutive positions for clarity), 
where we consider a clean~($W=0$) NN model with $N=6$, $L=12$, $V/t=50$, and open boundary conditions.  
Dashed lines indicate states with the same $N_{\mathrm{NN}}$. States with $2\leq N_{\mathrm{NN}} \leq 4$ show sub-blocks, formed by states with 
different cluster distribution, disconnected under unitary dynamics. 



\begin{figure}[tbp]
\includegraphics[width=0.9\columnwidth]{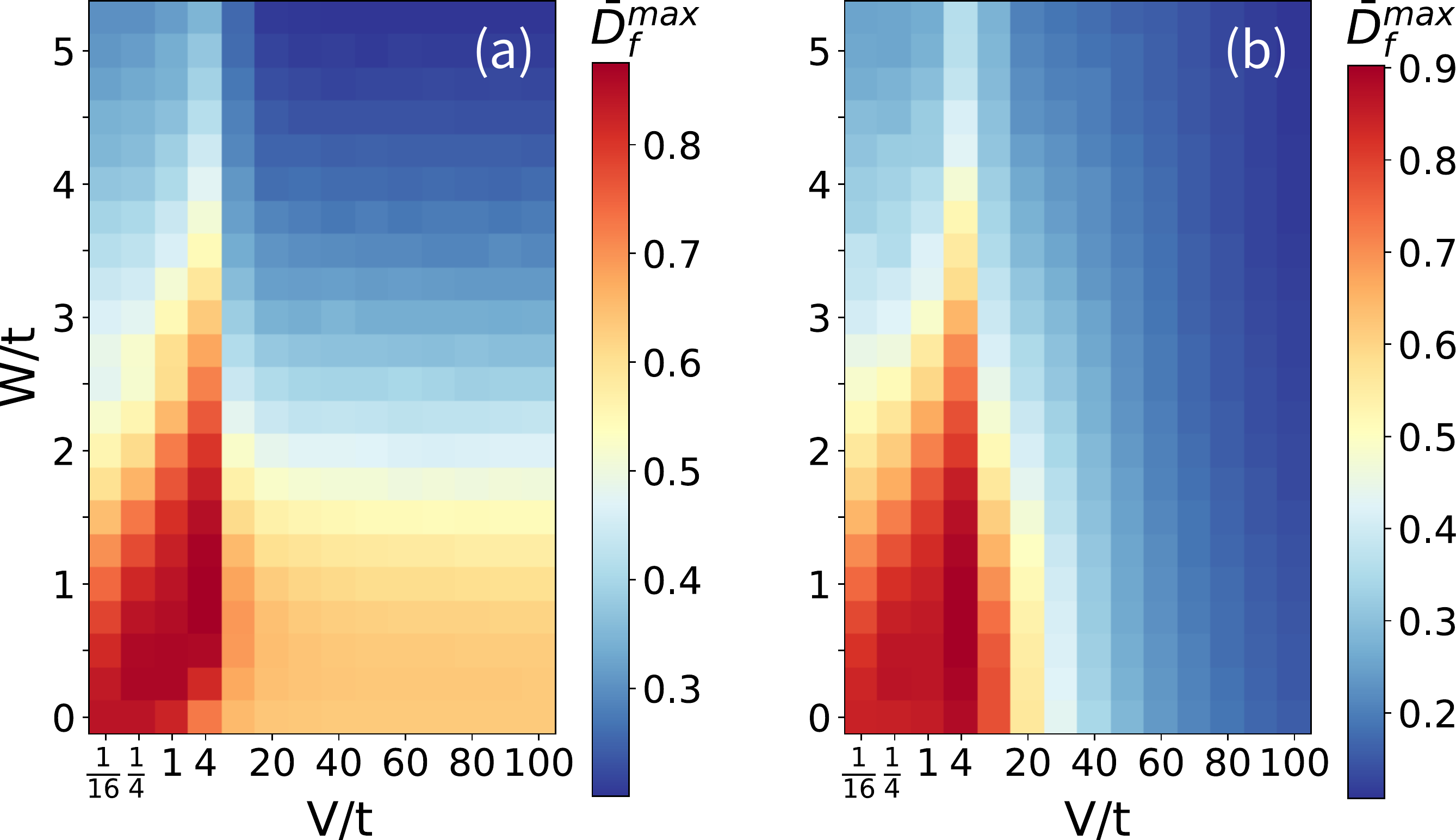} \centering
\caption{(Color online) $\bar D_f^{max}$ as a function of $V/t$ and $W/t$ for $N=8$ and $L=16$ for the NN model~(a) and the polar lattice gas~(b). 
Blue regions are regimes of strong fragmentation of the blocks of the clean NN model. 
Results obtained from exact diagonalization averaging over $1000$ different disorder realizations. 
The apparent abrupt change at $V/t\simeq 10$ is due to the reduction of the size $\Lambda_f$ of the block of maximal $\bar D_f$.}
\label{fig:2}
\end{figure}


Once fragmentation is determined in the clean NN model, further fragmentation either due to disorder or due to the $1/r^3$ dipolar tail 
may be analyzed by means of the inverse participation ratio $\mathrm{IPR}_f = \sum_{\alpha} |\psi_\alpha(f)|^{4}$ of a given 
Fock state $|f\rangle$. Starting an evolution with that state, $\mathrm{IPR}_f$ provides the long-time survival probability of the many-body state $|\psi(\tau)\rangle$
in the initial state, $|\langle f |\psi(\tau\to\infty) \rangle|^2$. Strong localization is hence characterized by $\mathrm{IPR}_f \sim 1$.
In the presence of disorder, $W>0$, localization results in the fragmentation of the NN blocks, which may be studied for $V/t>10$ 
by comparing $\mathrm{IPR}_f$ with the size $\Lambda_f$ of the Hilbert space block to which the particular Fock state 
belongs for the clean NN model. For $4<V/t<10$, blocks with fixed $N_{\mathrm{NN}}$ develop in the clean NN model, but the sub-block structure 
is not yet fully formed, and we hence set $\Lambda_f$ as the dimension of the block of states with fixed $N_{\mathrm{NN}}$. 
For $V/t<4$, no fragmentation occurs in the clean NN model and we hence fix 
$\Lambda_f$ as the dimension of the whole Hilbert space. 
Delocalized states (within the corresponding block of the clean NN model) are characterized by $\mathrm{IPR}_f\sim {\cal O}(\Lambda_f^{-1})$. 
We determine the fractal dimension, $D_f=-\ln(\mathrm{IPR}_f)/\ln(\Lambda_f)$~\cite{footnote-Mace, Mace2019}. $D_f\simeq 0$~($D_f\sim {\cal O}(1)$) implies localization~(delocalization)~\cite{footnote-localization}.  

Localization as a function of disorder vary from block to block (and even within the same block~\cite{DeTomasi2019}). 
For a given $V/t$ we determine the average $\bar D_f$ for each block, finding the block with the largest $\bar D_f^{max}$. 
For low filling factors, $\bar D_f^{max}$ corresponds, quite naturally, to the block with $N_{\mathrm{NN}}=0$. 
Note that when $\bar D_f^{max}\sim 0$ the whole spectrum localizes. 
Figure~\ref{fig:2}(a) shows $\bar D_f^{max}$ for a half-filled NN model~\cite{SM}. 
For growing $V/t$,  the region of delocalized states grows up to a maximum and then decreases due to the reduced cluster mobility, resulting in a re-entrant shape, in agreement with Ref.~\cite{BarLev2015}. 
However, disorder-induced fragmentation of the NN blocks does require a finite disorder strength even at $V=\infty$, due to the above-mentioned in-block resonant motion~\cite{footnote-Vinfinity}.



\begin{figure}[t]
\includegraphics[width=0.8\columnwidth]{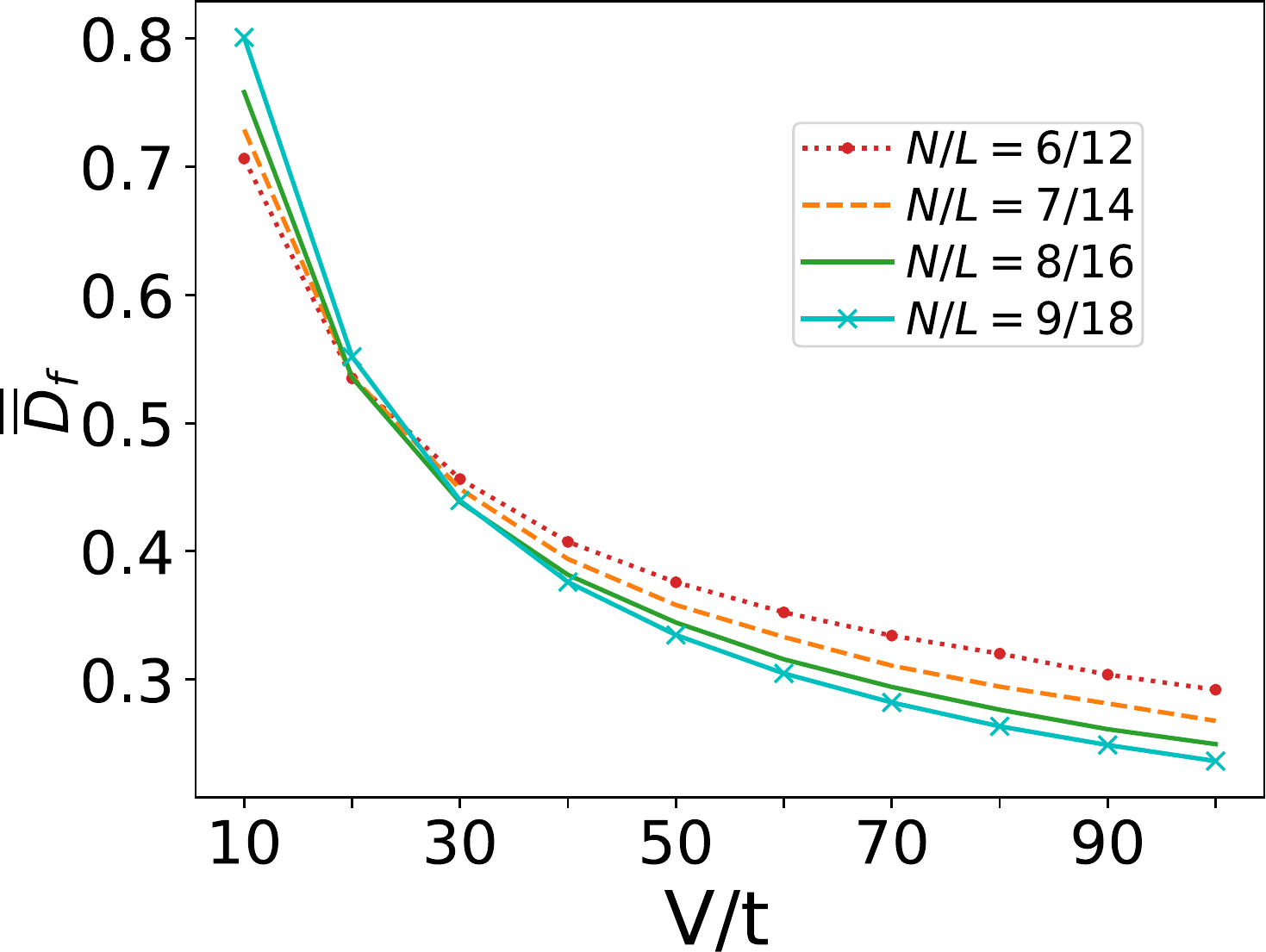} \centering
\caption{(Color online) Averaged fractal dimension $\bar {\bar D}_f$~(see text) as a function of $V/J$ for a half-filled clean polar lattice gas of different system sizes.  
For $V/t>20$, $\bar {\bar D}_f$ decreases with growing $L$ indicating a more pronounced shattering.
}
\label{fig:3}
\end{figure}




\paragraph{Polar lattice gas.--} 
Whereas in the NN model a growing $V/t$ just reenforces the conservation of $N_{\mathrm{NN}}$, in polar gases 
it leads to additional constraints, starting with the conservation of the number of next-to-NN bonds, 
$N_{\mathrm{NNN}}=\sum_j \langle n_j n_{j+2}\rangle$.  For $V\to\infty$ it seems intuitive that the conservation of the number of bonds at any distance 
leads to frozen dynamics even for $W=0$ for any initial condition~(there are however exceptions, as mentioned below). 
More interesting, however, is that dipolar interactions induce disorder-free quasi-localization~(see the discussion below) 
for values of $V/t$ well within experimental reach.

In polar lattice gases, the severe dynamical constraint induced by additional emerging conserved quantities results for sufficiently large $V/t$ in the shattering of the block structure of the NN model. 
As shown in Fig. ~\ref{fig:1}(b) for a clean half-filled case with $V/t=50$, the Hilbert space breaks down into a much finer structure compared to that of the clean NN model (Fig.~\ref{fig:1}(a)). 
Note that for $V/t=50$, interactions beyond next-to-NN are smaller than the bandwidth, $4t$, and hence, for half-filling the shattering of the NN blocks results from the mere additional emerging
conservation of $N_{\mathrm{NNN}}$.

As for the case of the disordered NN model, we may quantify the shattering of the NN blocks by means of the evaluation of 
the fractal dimension $D_f$ for the different Fock states, obtained again by comparing $\mathrm{IPR}_f$ with the size of the block evaluated for the clean NN model. In Fig.~\ref{fig:3} we depict 
for a clean polar lattice gas, for different system sizes, the average $\bar{\bar D}_f$ evaluated over the whole Fock basis, which 
provides a good quantitative estimation of the overall shattering of the NN blocks. The results show that clean polar lattice gases with $V/t \gtrsim 20$ are characterized by a strong shattering of the  
blocks of the clean NN model~\cite{SM}. The graph of $\bar D_f^{max}$~(Fig.~\ref{fig:2}(b)) is hence markedly different for $V/t>20$ compared to the NN model, with 
Hilbert-space shattering (and, as discussed below, localization) penetrating all the way to vanishingly small disorder.



\begin{figure*}[tbp]
\includegraphics[width=2\columnwidth]{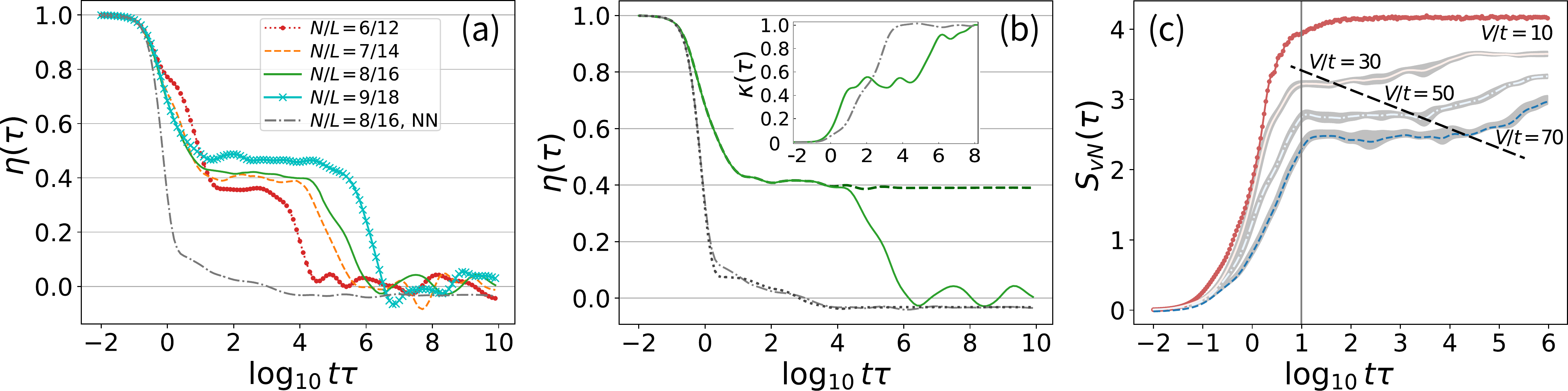} \centering
\caption{(Color online) (a) Time evolution of the homogeneity $\eta(\tau)$ for $V/t=50$ and periodic boundary conditions, for the NN model (with $N=8$ and $L=16$) and a polar gas (with different system sizes) 
The initial half-filled state is discussed in the text. In order to compare different system sizes we add or remove pairs $\fullcirc\, \emptycirc$ at the right of the state.
(b) Same as (a) for $N=8$ and $L=16$, comparing the clean case with that with a very small disorder $W/t=2\times 10^{-4}$~(averaged over $1000$ realizations). 
Solid green (dot-dashed grey) curves show the results with $W=0$ for the polar gas (NN model), and dashed green (dotted grey) those for the disordered polar gase (NN model).
In the inset, we depict $\kappa(\tau)$~(see text) for the NN model (dotted-dashed grey) and the polar gas (solid green), with  $N=8$ and $L=16$. 
(c) Entanglement entropy $S_{vN}$ evaluated for a partition of half of the polar lattice system, for $N=8$ and $L=16$ and different ratios $V/t$.}
\label{fig:4}
\end{figure*}




\paragraph{Dynamics.--} 
Whereas NN models are characterized by resonant motion within a Hilbert-space fragment, the emerging conservation of  $N_{\mathrm{NNN}}$ in a polar gas largely
prevents resonant dynamics. When the Hilbert space shatters, an initial Fock state can only connect resonantly to a limited number of Fock states in the same block, whereas 
the rest of the block can only be reached via virtual excursions into other Hilbert-space blocks in high order in $t/V\ll 1$.  
As a result, compared to the NN model, particle dynamics in the polar gas is typically dramatically slowed down for sufficiently large $V/t$. 

The latter is well illustrated by the evolution of the initial state 
$|\psi(\tau=0)\rangle = |\emptycirc\,\emptycirc\,\fullcirc\,\emptycirc\,\fullcirc\,\emptycirc\,\fullcirc\,\fullcirc\,\fullcirc\,\emptycirc\,\fullcirc\,\emptycirc\,\fullcirc\,\emptycirc\,\fullcirc\,\emptycirc\rangle$~(other initial 
states provide in general similar results). We employ exact time evolution of Eq.~\eqref{eq:H} and periodic boundary conditions to remove boundary effects. 
This initial half-filled state delocalizes in the NN model due to resonant hops, which break the central trimer into two dimers: 
$|\emptycirc\,\emptycirc\,\fullcirc\,\emptycirc\,\fullcirc\,\emptycirc\,\fullcirc\,\fullcirc\,\emptycirc\,\fullcirc\,\fullcirc\,\emptycirc\,\fullcirc\,\emptycirc\,\fullcirc\,\emptycirc\rangle$
and then delocalize each dimer
, e.g.  $|\emptycirc\,\emptycirc\,\fullcirc\,\emptycirc\,\fullcirc\,\fullcirc\,\emptycirc\,\fullcirc\,\emptycirc\,\fullcirc\,\fullcirc\,\emptycirc\,\fullcirc\,\emptycirc\,\fullcirc\,\emptycirc\rangle$. 
All these processes remain resonant in the NN model even for $V=\infty$. In contrast, a sufficiently large dipolar interaction, renders 
the breaking of the initial trimer non-resonant, since it does not preserve $N_{\mathrm{NNN}}$. Moreover, the formation of beyond-NN clusters further hinders the particle dynamics.

Analogous to MBL experiments based on the evolution of density waves~\cite{Schreiber2015, Scherg2020}, and similar to recent trap ion experiments~\cite{Morong2021}, 
we define the homogeneity parameter as  $\eta(\tau)=\frac{N_0(\tau)/L_0 - N/L}{1-N/L}$,  where
$N_0(\tau)=\sum_{j\in f_0} \langle n_j (\tau) \rangle$ is the number of particles in the set $f_0$ of $L_0$ initially occupied sites. Homogeneization of the on-site populations results in $\eta(\tau)\to 0$. 
We depict in Fig.~\ref{fig:4}(a) $\eta(\tau)$ for $V/t=50$. Homogeneization is quickly reached for the NN model at $\tau\sim 1/t$~(tiny residual 
values are due to finite size), whereas for a polar gas, $\eta$ plateaus at a large value, indicating a long-lived memory of initial conditions. We have checked that 
for this example the plateau is already evident for $V/t > 20$. 

A longer-time evolution reveals eventual delocalization due to a very weak coupling between Fock states belonging to the same small block of the shattered Hilbert space. This coupling  
results from the above-mentioned higher-order virtual excursions to other blocks. 
For large-enough $V/t$, many such virtual excursions are necessary, and hence the coupling between Fock states becomes exponentially small in $t/V$. 
This quasi-localization within the block $B$ (of size $\Omega_f$) of the shattered Hilbert space to which $|\psi(\tau = 0)\rangle$ belongs 
is well visualized by monitoring $|\psi(\tau>0)\rangle=\sum_{f\in B} \psi_f(\tau) |f\rangle$~\cite{footnote-small-blocks}, and determining the participation ratio $\mathrm{PR}(\tau) = \left (\sum_{f\in B} |\psi_f(\tau)|^4 \right ) ^{-1}$. 
In the inset of Fig.~\ref{fig:4}(b) we depict $\kappa(\tau)=\mathrm{PR}(\tau)/\mathrm{PR}(\infty)$ showing that the long-lived memory of initial conditions observed in $\eta(\tau)$
results from the fact that only a limited fraction of the Hilbert-space block is effectively reached during the plateau time.
 
The two-stage dynamics is evident in the evolution of the entanglement entropy, $S_{vN}$,  calculated from a partition of the system to one half~(Fig.~\ref{fig:4}(c)). 
In contrast to standard MBL, the lack of local integrals of motion results in the absence of logarithmic growth 
of $S_{vN}$, which plateaus during the localization, and only grows due to the eventual delocalization at finite $V/t$.
The time of the on-set of the second stage scales exponentially with $V/t$, being observable for $N=8$ and $L=16$ for $V/t\simeq 30$ but prohibitively long for typical experiments for $V/t>50$. 
Moreover, our analysis of different system sizes~(Fig.~\ref{fig:4}(a)) shows that the delocalization time also scales exponentially with the system size, since even more intricate virtual excursions are needed to 
connect different Fock states in the block. Note also that the high-order excursions responsible for the eventual delocalization are cancelled even by vanishingly small disorders, as shown in 
Fig.~\ref{fig:4}(b), whereas the same tiny disorder has a negligible effect for the NN model~\cite{SM}. 
 
Finally, note that as for other systems with kinetic constraints, particle dynamics strongly depends on the particular initial condition. 
Although the lack of a general resonant motion mechanism will slow down and quasi-localize typical states, some particular states may remain delocalized 
even for very large $V/t$. This is the case of a density wave with a single domain wall, e.g.
$|\dots \fullcirc\,\emptycirc\,\fullcirc\,\emptycirc\,\fullcirc\,\emptycirc\,\fullcirc\,\emptycirc\,\emptycirc\,\fullcirc\,\emptycirc\,\fullcirc\,\emptycirc\,\fullcirc\,\emptycirc\,\fullcirc\dots\rangle$. 
With periodic boundary conditions, the wall moves resonantly while preserving the interaction energy to all neighbors, delocalizing
for any arbitrary $V$. Note, however, that for open boundary conditions the boundaries induce, due to the dipolar tail, an effective confinement for the domain 
wall, preventing it to reach a distance $(V/t)^{1/3}$ from the lattice edges~\cite{SM}. This is yet another localization mechanism induced by the dipolar tail that may be relevant in experiments.



\paragraph{Conclusions.--} Emerging dynamical constraints induced by 
the dipolar $1/r^3$ tail lead to Hilbert-space shattering, which at half-filling occurs for $V/t\gtrsim 20$. 
Moreover,  these constraints disrupt the resonant transport characteristic of NN models, 
resulting in a dramatic slow-down of the particle dynamics and eventual disorder-free localization.
Although we have focused on the dynamics once shattering develops, a significant 
slow-down also occurs within the not yet shattered NN blocks even for smaller $V/t$ ratios. 
This interesting dynamics will be the focus of a forthcoming work.

For magnetic atoms, recent experiments have achieved $V/t\simeq 3$~\cite{Baier2016}, but the use of Feshbach molecules of lanthanide atoms~\cite{Frisch2015} 
and/or subwavelength~\cite{Yi2008, Nascimbene2015} or UV lattices may significantly boost the $V/t$ ratio. 
For example, for $^{164}$Dy in an UV lattice with $180$nm spacing and depth of $23$ recoil energies, $|V|/t\simeq 30$, with $t/\hbar\simeq 93$s$^{-1}$. 
Disorder-free localization could then be probed in few seconds, well within experimental lifetimes. 
Polar molecules offer exciting possibilities for large $V/t$ even without the need of a special lattice, due to their
much stronger dipolar interaction, orders of magnitude larger than that of magnetic atoms~\cite{Moses2017}. 
Our work hence shows that the study of the interplay between disorder- and 
interaction-induced localization is well within reach of future experiments on polar lattice gases. Moreover, our results 
have a more general applicability, being potentially relevant for other disorder-free systems with more general long-range interactions, in particular 
trapped ions~\cite{Morong2021}, where intriguing localization properties may result from the interplay between power-law exchange and Ising terms.



We acknowledge support by the Deutsche Forschungsgemeinschaft (DFG, German Research Foundation) under the project SA 1031/11, 
the SFB 1227 ``DQ-mat'', project A04, and under Germany's Excellence Strategy -- EXC-2123 QuantumFrontiers -- 390837967.



\begin{thebibliography}{99}
\bibitem{footnote-hardcore} The hard-core constraint requires large-enough on-site interactions, which may demand the use of Feshbach resonances.
\bibitem{footnote-localization} This definition is only problematic for large $V/t$ for some states, that for $W=0$ and $V\to\infty$ remain frozen 
in the clean NN model~(in particular states without singlons), which have by definition 
$\Lambda_f=1$. However, the definition is unproblematic for the bulk of states of the NN model with a finite density of moving particles, which for $W=0$ 
are characterized by blocks whose $\Lambda_f$ grows with the system size.
\bibitem{footnote-small-blocks} The Fock states belonging to the block $B$ are those which are eventually connected to the initial states. We determine them by ordering the Fock states $|f\rangle$ with a decreasing value of 
 $|\langle f | \psi(\tau \to\infty)\rangle |^2 $, where $| \psi(\tau \to\infty)\rangle$ is the asymptotic time-evolved state, and then summing them until reaching  $\sum_{f \in B} |\langle f | \psi(\tau \to\infty)\rangle |^2 = 0.95$.
\end{thebibliography}
\end{document}